\documentstyle[twoside,fleqn,espcrc2,epsfig]{article}

\def\gtwid{\mathrel{\raise.3ex\hbox{$>$\kern-.75em\lower1ex\hbox{$\sim$}}}}
\def\ltwid{\mathrel{\raise.3ex\hbox{$<$\kern-.75em\lower1ex\hbox{$\sim$}}}}

\newcommand{\AmS}{{\protect\the\textfont2
  A\kern-.1667em\lower.5ex\hbox{M}\kern-.125emS}}

\title{Axion Searches}

\author{Pierre Sikivie\address{Department of Physics,
        University of Florida, \\
        Gainesville, FL 32611, USA}%
        \thanks{This work is supported in part by the U.S. Department 
                of Energy under contract DE-FG05-86ER40272.}}

\begin{document}

\begin{abstract}
The strong CP problem and its resolution through the existence of an
axion are briefly reviewed.  The constraints on the axion from 
accelerator searches, from the evolution of red giants and from 
supernova SN1987a combine to require $m_a < 3 \cdot 10^{-3}$ eV, 
where $m_a$ is the axion mass.  On the other hand, the constraint 
that axions do not overclose the universe implies $m_a \gtwid 10^{-6}$ eV.  
If $m_a \sim 10^{-5}$ eV, axions contribute significantly to the 
cosmological energy density in the form of cold dark matter.  Dark 
matter axions can be detected by resonant conversion to microwave 
photons in a cavity permeated by a static magnetic field and tuned 
to the axion mass.  Experiments using this effect are described, as 
well as several other types of axion searches.
\end{abstract}

\maketitle

\section{Introduction}

The axion was postulated nearly two decades ago \cite{arev} to explain 
why the strong interactions conserve $P$ and $CP$ in spite of the fact 
that the weak interactions violate those symmetries.  Consider the 
Lagrangian of QCD:
\begin{eqnarray}
{\cal L}_{QCD} = -{1\over 4} G^a_{\mu\nu} G^{a\mu\nu} +\sum_{j=1}^n
\left[ \overline q_j \gamma^\mu i D_\mu q_j \right. \nonumber\\
- \left. (m_j q_{Lj}^\dagger q_{Rj} + \hbox{h.c.})\right]
+ {\theta g^2\over 32\pi^2} G^a_{\mu\nu} \tilde G^{a\mu\nu} ~~\ .
\end{eqnarray}
The last term is a 4-divergence and hence does not contribute in 
perturbation theory.  That term does however contribute through 
non-perturbative effects \cite{thft} associated with QCD instantons 
\cite{inst}.  Such effects can make the physics of QCD depend upon the 
value of $\theta$.  Using the Adler-Bell-Jackiw anomaly \cite{abj}, one 
can show that $\theta$ dependence must be present if none of the current 
quark masses vanishes.  Indeed otherwise QCD would have a $U_A(1)$ 
symmetry and would predict the mass of the $\eta'$ pseudo-scalar meson 
to be less than $\sqrt{3} m_\pi \approx 240$~MeV \cite{SW}, contrary to 
observation.  Using the anomaly, one can further show that QCD depends 
upon $\theta$ only through the combination of parameters:
\begin{equation}
\overline \theta =\theta -\hbox{arg} (m_1, m_2, \ldots m_n) ~~~\ .
\end{equation}
If $\overline \theta \neq 0$, QCD violates $P$ and $CP$.  The absence of
$P$ and $CP$ violations in the strong interactions therefore places an
upper limit upon $\overline\theta$.  The best constraint follows from 
the experimental bound \cite{ned} on the neutron electric dipole moment 
which yields: $\overline\theta < 10^{-9}$.

The question then is:  why is $\overline\theta$ so small?  In the 
Standard Model of particle interactions, the quark masses originate 
in the electroweak sector which violates $P$ and $CP$.  There 
is no reason why the overall phase of the quark mass matrix should
exactly match the value of $\theta$ from the QCD sector to yield
$\overline \theta < 10^{-9}$. In particular, if $CP$ violation is 
introduced in the manner of Kobayashi and Maskawa \cite{KM}, the Yukawa 
couplings that give masses to the quarks are arbitrary complex numbers 
and hence arg~det~$m_q$ and $\overline\theta$ are expected to be of order
one.

The problem why $\overline\theta < 10^{-9}$ is usually referred to as the 
``strong $CP$ problem''.  The existence of an axion solves this problem
in a simple manner which is rich in implications for experiment, for 
astrophysics and for cosmology.  There are two alternative solutions 
however. The first alternative is to set $m_u = 0$.  This removes the 
$\theta$-dependence of QCD and thus solves the strong $CP$ problem.  
The well-known calculation of the pseudo-scalar meson masses in lowest 
order of chiral perturbation theory yields $m_u \simeq 4$ MeV, which is 
incompatible with $m_u = 0$.  This calculation also predicts the successful 
Gell-Mann - Okubo relation among the pseudo-scalar masses squared.  It is 
possible to have $m_u = 0$ by invoking second order effects \cite{Georgi81}.
This a reasonable proposition because $m_s$ happens to be of order the
QCD scale.  However, when second order effects are included \cite{Gasser}, 
the Gell-Mann - Okubo relation is in general violated.  Thus the price for 
having $m_u = 0$ through higher order effects is that the Gell-Mann - Okubo 
relation becomes an accident.  The second alternative solution to the strong 
CP problem is to assume that $CP$ and/or $P$ is spontaneously broken but 
is otherwise a good symmetry.  In this case, $\overline\theta$ is calculable 
and may be arranged to be small \cite {CPsb}.  

In addition, it is worth emphasizing that the strong $CP$ problem need not 
be solved in the low energy theory.  Indeed, as Ellis and Gaillard \cite{EG} 
pointed out, if in the standard model $\overline\theta=0$ near the Planck 
scale then $\overline\theta \ll 10^{-9}$ at the QCD scale.  

Peccei and Quinn \cite{PQ} proposed to solve the strong $CP$ problem by 
postulating the existence of a global $U_{PQ}(1)$ quasi-symmetry.  $U_{PQ}(1)$ 
must be a symmetry of the theory at the classical (i.e., at the Lagrangian) 
level, it must be broken explicitly by those non-perturbative effects that 
make the physics of QCD depend upon $\theta$, and finally it must be 
spontaneously broken.  The axion \cite{WW} is the quasi-Nambu-Goldstone boson 
associated with the spontaneous breakdown of $U_{PQ}(1)$.  One can show that, 
if a $U_{PQ}(1)$ quasi-symmetry is present, then
\begin{equation}
\overline\theta = \theta - arg (m_1 \ldots m_n) - {a(x)\over f_a}\, ,
\end{equation}
where $a(x)$ is the axion field and $f_a = v/N$ is called the axion decay 
constant.  $v$ is the vacuum expectation value which spontaneously
breaks $U_{PQ}(1)$ and $N$ is an integer which expresses the color anomaly 
of $U_{PQ}(1)$.  Axion models have $N$ degenerate vacua \cite{adw}.  The 
non-perturbative effects that make QCD depend upon $\overline\theta$ 
produce an effective potential $V(\overline\theta)$ whose minimum is  
at $\overline\theta =0$. \cite{VW}\ \ Thus, by postulating an axion, 
$\overline\theta$ is allowed to relax to zero dynamically and the strong 
$CP$ problem is solved.

The properties of the axion can be derived using the methods of current
algebra \cite{curr}.  The axion mass is given in terms of $f_a$ by
\begin{equation}
m_a\simeq 0.6~eV~{10^7 GeV\over f_a}\, .
\end{equation}
All the axion couplings are inversely proportional to $f_a$.  Of particular 
interest here is the axion coupling to two photons:
\begin{equation}
{\cal L}_{a\gamma\gamma} = -g_\gamma {\alpha\over \pi} {a(x)\over f_a}
\vec E \cdot\vec B
\end{equation}
where $\vec E$ and $\vec B$ are the electric and magnetic fields, 
$\alpha$ is the fine structure constant, and $g_\gamma$ is a model-dependent 
coefficient of order one.  $g_\gamma=0.36$ in the DFSZ model \cite{DFSZ}
whereas $g_\gamma=-0.97$ in the KSVZ model \cite{KSVZ}.  The coupling 
of the axion to a spin 1/2 fermion $f$ has the form:
\begin{equation}
{\cal L}_{a \overline f f} = i g_f {m_f \over v} a \overline f \gamma_5 f
\label{cf}
\end{equation}
where $g_f$ is a model-dependent coefficient of order one.  In the 
KSVZ model the coupling to electrons is zero at tree level.  Models 
with this property are called 'hadronic'.

\section{Constraints from laboratory searches and astrophysics}

The searches for the axion in high energy and nuclear physics experiments 
are discussed in the reviews by Kim \cite{arev} and by Peccei \cite{arev}.  
A complete list of experiments can be found in the Review of Particle 
Physics \cite{RPP}.  If the axion is heavier than 1 MeV and decays 
quickly into $e^+e^-$ (lifetime of order $10^{-11}$ sec or less), then 
it is ruled out by negative searches for rare particle decays such as 
$\pi^+ \rightarrow a(e^+ e^-) e^+ \nu_e$ \cite{SIN}.  The rate of this 
reaction follows simply from the mixing of $a$ and $\pi^0$ and the known 
decay $\pi^+ \rightarrow \pi^0 e^+ \nu_e$, and hence it is almost 
completely model-independent \cite{Suz}.  Alternatively, the axion is 
long lived ($10^{-11}$ sec or more).  In this case it is severely constrained 
by negative searches in beam dumps.  In a beam dump, axions are produced 
by many different processes involving independent couplings, such as 
$a - \pi^0$, $a - \eta$ and $a - \eta^\prime$ mixing, the axion couplings 
to two photons and to two gluons, and the couplings to quarks and gluons.  
It is difficult to calculate these processes with precision at all energies.  
On the other hand, the many processes add up incoherently and they can not 
all vanish.  Thus one can give a reliable estimate for the total production 
($p + N \rightarrow a + X$, or $e + N \rightarrow a + X$) and interaction 
($a + N \rightarrow X$) cross-sections.  Many such searches \cite{RPP} 
have been carried out.  They rule out axions with mass larger than about 
50 keV. 

The astrophysical contraints on the axion are described in detail in 
the reviews by Turner \cite{arev} and Raffelt \cite{arev}.  Axions are
emitted by stars in a variety of processes such as Compton-like scattering
($\gamma + e \rightarrow a + e$), axion bremstrahlung ($e + N \rightarrow
N + e + a$) and the Primakoff process ($\gamma + N \rightarrow N + a$).
The rate at which a star burns its nuclear fuel is limited by the rate at 
which it can loose the energy produced.  Emission of light weakly coupled 
bosons, such as the axion, allows a star to radiate energy efficiently 
because such particles can escape the star at once (without 
rescattering) whereas photons are emitted only from the stellar 
surface \cite{Sato}.  Thus the existence of an axion accelerates 
stellar evolution, which may be inconsistent with observation.  The 
longevity of red giants rules out the mass range 
200 keV $\gtwid m_a \gtwid 0.5$ eV \cite{Dicus,Raff87} for 
hadronic axions.  Above 200 keV the axion is too heavy to be copiously 
emitted in the thermal processes taking in red giants, whereas below 
0.5 eV it is too weakly coupled.  For axions with a large coupling to 
electrons [$g_e = 0(1)$ in Eq. \ref{cf}] the ruled out range can be 
extended to 200 keV $\gtwid m_a \gtwid 10^{-2}$ eV because axion 
emission through the Compton-like process $\gamma + e \rightarrow a + e$
cools the helium core to such an extent as to prevent the onset of 
helium burning \cite{Schramm}.  

Finally the range 2 eV $\gtwid m_a \gtwid 3 \cdot 10^{-3}$ eV is ruled 
out by Supernova 1987a \cite{1987a}.  The constraint follows from the 
fact that the duration of the associated neutrino events in the large 
underground proton decay detectors \cite{ugd} is consistent with 
theoretical expectations based on the premise that the collapsed 
supernova core cools by emission of neutrinos.  If the axion mass 
is in the above-mentioned range, the core cools instead by axion 
emission and the neutrino burst is excessively shorthened.  The 
supernova constraint is quite axion model-independent because the 
axions are emitted by axion bremstrahlung in nucleon-nucleon scattering 
($N + N \rightarrow N + N + a$) and the relevant couplings follow
simply from the mixing of the axion with the $\pi^0$ which is a
general feature of axion models.

When the limits from laboratory searches are combined with the 
astrophysical contraints, all of the axion mass range down to 
approximately $3 \cdot 10^{-3}$ eV is ruled out.

\section{Axion cosmology}

The implications of the existence of an axion for the history of the
early universe may be briefly described as follows.  At a temperature of
order $v$, a phase transition occurs in which the $U_{PQ}(1)$ symmetry  
becomes spontaneously broken.  This is called the PQ phase transition. 
At these temperatures, the non-perturbative QCD effects which produce the 
effective potential $V(\overline\theta)$ are suppressed \cite{GPY}, the 
axion is massless and all values of $\langle a(x)\rangle$ are equally 
likely.  Axion strings appear as topological defects.  One must
distinguish two cases: 1) inflation occurs with reheat temperature 
higher than the PQ transition temperature (equivalently, for our 
purposes, inflation does not occur at all) or 2) inflation occurs 
with reheat temperature less than the PQ transition temperature.  In 
case 2 the axion field gets homogenized by inflation and the axion 
strings are 'blown' away.  

When the temperature approaches the QCD scale, the potential 
$V(\overline\theta)$ turns on and the axion acquires mass.  There is 
a critical time, defined by $m_a(t_1)t_1 = 1$, when the axion field 
starts to oscillate in response to the turn-on of the axion mass.  
The corresponding temperature $T_1 \simeq 1$ GeV \cite{ac}.  The 
initial amplitude of this oscillation corresponds to how far from 
zero the axion field is when the axion mass turns on.  The axion 
field oscillations do not dissipate into other forms of energy and 
hence contribute to the cosmological energy density today \cite{ac}.  
This contribution is called of `vacuum realignment'.  It is further 
described below.  Note that the vacuum realignment contribution may 
be accidentally suppressed in case 2 because the homogenized axion 
field happens to lie close to zero.

In case 1 the axion strings radiate axions \cite{rd,Har} from the 
time of the PQ transition till $t_1$ when the axion mass turns on.   At
$t_1$ each string becomes the boundary of $N$ domain walls.  If $N=1$, 
the network of walls bounded by strings is unstable \cite{Ev,Paris} and
decays away.  If $N>1$ there is a domain wall problem \cite{adw} because
axion domain walls end up dominating the energy density, resulting in a
universe very different from the one observed today.  There is a way 
to avoid this problem by introducing an interaction which slightly 
lowers one of the $N$ vacua with respect to the others.  In that
case, the lowest vacuum takes over after some time and the domain walls
disappear.  There is little room in parameter space for that to happen
and we will not consider this possibility further here.  A detailed
discussion is given in Ref.~\cite{axwall}.  Henceforth, we assume $N=1$.

In case 1 there are three contributions to the axion cosmological
energy density.  One contribution \cite{rd,Har,thA,Hag,Shel,Yam,us}
is from axions that were radiated by axion strings before $t_1$.  A 
second contribution is from axions that were produced in the decay 
of walls bounded by strings after $t_1$ \cite{Hag,Ly,Nag,axwall}.  A 
third contribution is from vacuum realignment \cite{ac}. 

Let me briefly indicate how the vacuum alignment contribution is 
evaluated.  Before time $t_1$, the axion field did not oscillate 
even once.  Soon after $t_1$, the axion mass is assumed to change 
sufficiently slowly that the total number of axions in the 
oscillations of the axion field is an adiabatic invariant.  The 
number density of axions at time $t_1$ is
\begin{equation}
n_a(t_1)\simeq {1\over 2} m_a(t_1) \langle a^2(t_1)\rangle \simeq
\pi f_a^2 {1\over t_1}
\label{nat1}
\end{equation}
where $f_a$ is the axion decay constant introduced earlier.  
In Eq.~(\ref{nat1}), we used the fact that the axion field $a(x)$ 
is approximately homogeneous on the horizon scale $t_1$.  Wiggles 
in $a(x)$ which entered the horizon long before $t_1$ have been 
red-shifted away \cite{Vil}.  We also used the fact that the initial 
departure of $a(x)$ from the nearest minimum is of order $f_a$.  The 
axions of Eq.~(\ref{nat1}) are decoupled and non-relativistic.  
Assuming that the ratio of the axion number density to the entropy 
density is constant from time $t_1$ till today, one finds \cite{ac}
\begin{equation}
\Omega_a \simeq \left({0.6~10^{-5}\hbox{\ eV}\over m_a}\right)^{7\over 6}
\left({200\hbox{\ MeV}\over \Lambda_{QCD}}\right)^{3\over 4} h^{-2}
\label{oma}
\end{equation}
for the ratio of the axion energy density to the critical density
for closing the universe.  $h$ is the present Hubble rate in units 
of 100 km/s.Mpc.  The requirement that axions do not overclose the 
universe implies the constraint $m_a \gtwid 6 \cdot 10^{-6}$~ eV.

The contribution from axion string decay has been debated over the 
years.  The main issue is the energy spectrum of axions radiated 
by axion strings.  Battye and Shellard \cite{Shel} have carried out 
computer simulations of bent strings (i.e. of wiggles on otherwise 
straight strings) and have concluded that the contribution from 
string decay is approximately ten times larger than that from vacuum 
realignment, implying a bound on the axion mass approximately then 
times more severe, say $m_a \gtwid 6 \cdot 10^{-5}~eV$ instead of 
$m_a \gtwid 6 \cdot 10^{-6}~eV$.  My collaborators and I have done
simulations of bent strings \cite{Hag}, of circular string loops 
\cite{Hag,us} and non-circular string loops \cite{us}.  We conclude 
that the string decay contribution is of the same order of magnitude 
than that from vacuum realignment.  Recently, Yamaguchi, Kawasaki and 
Yokoyama \cite{Yam} have done computer simulations of a network of 
strings in an expanding universe, and concluded that the contribution 
from string decay is approximately three times that of vacuum 
realignment.  The contribution from wall decay has been discussed 
in detail in ref.~\cite{axwall}.  It is probably subdominant compared 
to the vacuum realignment and string decay constributions.  

It should be emphasized that there are many sources of uncertainty
in the cosmological axion energy density aside from the uncertainty 
about the constribution from string decay.  The axion energy density may
be diluted by the entropy release from heavy particles which decouple
before the QCD epoch but decay afterwards \cite{ST}, or by the entropy
release associated with a first order QCD phase transition.  On the other
hand, if the QCD phase transition is first order \cite{pt}, an abrupt
change of the axion mass at the transition may increase $\Omega_a$.  If
inflation occurs with reheat temperature less than $T_{PQ}$, there   
may be an accidental suppression of $\Omega_a$ because the homogenized
axion field happens to lie close to a $CP$ conserving minimum.  Because
the RHS of Eq.~(7) is multiplied in this case by a factor of order the
square of the initial vacuum misalignment angle ${a(t_1)\over v}N$   
which is randomly chosen between $-\pi$ and $+\pi$, the probability  
that $\Omega_a$ is suppressed by a factor $x$ is of order $\sqrt{x}$. 
This rule cannot be extended to arbitrarily small $x$ however because  
quantum mechanical fluctuations in the axion field during the epoch of
inflation do not allow the suppression to be perfect \cite{inflax}.

The axions produced when the axion mass turns on during the QCD phase 
transition are cold dark matter (CDM) because they are non-relativistic 
from the moment of their first appearance at 1~GeV temperature.  Studies 
of large scale
\begin{figure}[top]
\vspace{2mm}
\epsfxsize=50mm
\centerline{\epsfbox{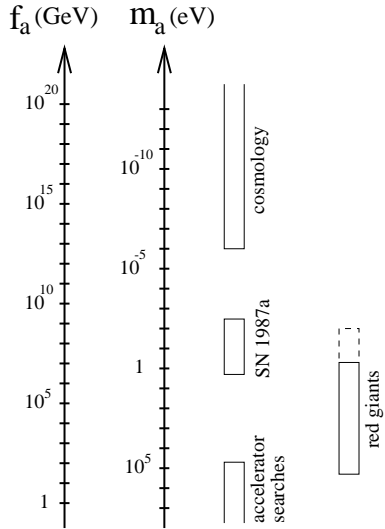}}
\vspace{-6mm}
\caption{\small{Ranges of axion mass $m_a$, or equivalently axion 
decay constant $f_a$, which have been ruled out by accelerator 
searches, the evolution of red giants, the supernova SN1987a, 
and finally the axion cosmological energy density.}}
\label{fig1}
\vspace{-3mm}
\end{figure}
structure formation support the view that the dominant fraction
of dark matter is CDM.  Any form of CDM necessarily contributes to
galactic halos by falling into the gravitational wells of galaxies.  
Hence, there is excellent motivation to look for axions as 
constituent particles of our galactic halo.

Finally, let's mention that there is a particular kind of clumpiness
\cite{amc,axwall} which affects axion dark matter if there is no inflation
after the Peccei-Quinn phase transition.  This is due to the fact that the
dark matter axions are inhomogeneous with $\delta \rho / \rho \sim 1$ over
the horizon scale at temperature $T_1 \simeq$ 1 GeV, when they 
are produced at the start of the QCD phase-transition, combined  
with the fact that their velocities are so small that they do not
erase these inhomogeneities by free-streaming before the time $t_{eq}$
of equality between the matter and radiation energy densities when
matter perturbations can start to grow.  These particular inhomogeneities
in the axion dark matter are in the non-linear regime immediately after
time $t_{eq}$ and thus form clumps, called `axion mini-clusters'
\cite{amc}.  They have mass $M_{mc} \simeq 10^{-13} M_\odot$ and
size $l_{mc} \simeq 10^{13}$ cm.

The various constraints on the axion, from accelerator seaches,  
astrophysics and cosmology are summarized in Fig.~\ref{fig1}.

\section{The cavity detector of galactic halo axions}

An electromagnetic cavity permeated by a strong static magnetic field 
can be used to detect galactic halo axions \cite{ps}.  The relevant 
coupling is given in Eq.~(5). Galactic halo axions have velocities 
$\beta$ of order $10^{-3}$ and hence their energies 
$E_a=m_a+{1\over 2} m_a\beta^2$ have a spread of order $10^{-6}$ 
above the axion mass.  When the frequency $\omega=2\pi f$ of a 
cavity mode equals $m_a$, galactic halo axions convert resonantly 
into quanta of excitation (photons) of that cavity mode.  The power 
from axion $\to$ photon conversion on resonance is found to 
be \cite{ps,kal}:
\begin{eqnarray}
P=\left ({\alpha\over\pi} {g_\gamma\over f_a}\right )^2 V\, B_0^2 
\rho_a C {1\over m_a} \hbox{Min}(Q_L,Q_a)~~~~\nonumber\\
= 0.5\; 10^{-26} \hbox{Watt}\left( {V\over 500\hbox{\ liter}}\right) 
\left({B_0\over 7\hbox{\ Tesla}}\right)^2 \nonumber\\
\cdot~C \left({g_\gamma \over 0.36}\right)^2  
\left({\rho_a\over {1\over 2} \cdot 10^{-24}
{{\scriptstyle g_r}\over \hbox{\scriptsize cm}^3}}\right) \nonumber\\ 
\cdot~\left({m_a\over 2\pi (\hbox{GHz})}\right)\hbox{Min}(Q_L,Q_a)
\end{eqnarray}
where $V$ is the volume of the cavity, $B_0$ is the magnetic field 
strength, $Q_L$ is its loaded quality factor, $Q_a=10^6$ is the 
`quality factor' of the galactic halo axion signal (i.e. the ratio of 
their energy to their energy spread), $\rho_a$ is the density of 
galactic halo axions on Earth, and $C$ is a mode dependent form factor
given by
\begin{equation}
C = {\left| \int_V d^3 x \vec E_\omega \cdot \vec B_0\right|^2
\over B_0^2 V \int_V d^3x \epsilon |\vec E_\omega|^2}  \, 
\end{equation}
where $\vec B_0(\vec x)$ is the static magnetic field,
$\vec E_\omega(\vec x) e^{i\omega t}$ is the oscillating electric
field and $\epsilon$ is the dielectric constant.

Because the axion mass is only known in order of magnitude at
best, the cavity must be tunable and a large range of frequencies
must be explored seeking a signal.  The cavity can be tuned by
moving a dielectric rod or metal post inside it.  Using Eq.~(8), one 
finds the scanning rate to perform a search with signal to noise 
ratio $s/n$:
\begin{eqnarray}
{df\over dt} &=& {12 \hbox{GHz}\over \hbox{year}} \left({4n\over s}
\right)^2 \left({V\over 500\hbox{\ liter}}\right)^2
\left( {B_0\over 7\hbox{\ Tesla}}\right)^4 \nonumber\\
&\cdot&~C^2\left({g_\gamma \over 0.36}\right)^4 
\left({\rho_a\over {1\over 2}\cdot 10^{-24} 
{{\scriptstyle gr}\over \hbox{\scriptsize cm}^3}}\right)^2 
\left({3K\over T_n}\right)^2 \nonumber\\
&\cdot&~\left({f\over \hbox{GHz}}\right)^2 {Q_L\over Q_a}~~ ,
\end{eqnarray}
where $T_n$ is the sum of the physical temperature of the cavity plus 
the electronic noise temperature of the microwave receiver that detects 
the photons from $a \to \gamma$ conversion.  Eq.~(11) assumes that 
$Q_L < Q_a$ and that some strategies have been followed which optimize 
the search rate.  The best quality factors attainable at present, using 
oxygen free copper, are of order $10^5$ in the GHz range. 

\begin{figure}[top]
\vspace{2mm} 
\epsfxsize=75mm
\centerline{\epsfbox{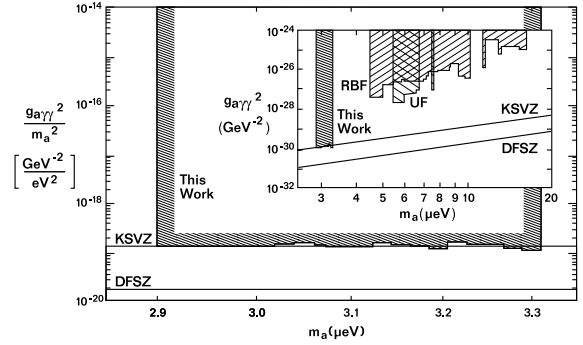}}
\vspace{-6mm}
\caption{\small{Axion couplings and masses excluded by the Large Scale US
Dark Matter Axion Search at LLNL.  Also shown are the KSVZ and DFSZ model 
predictions.  Indicated on the insert are the regions excluded by the 
pilot experiments at Brookhaven National Laboratory (RBF) and the 
University of Florida (UF).  All results are scaled to 
$\rho_a = 7.5~10^{-25}~$g/cm$^3$.}}
\label{fig2}
\vspace{-3mm}
\end{figure}

Eq. (11) shows that a galactic halo search with the required
sensitivity is feasible with presently available technology,
provided the form factor $C$ can be kept at values of order one
for a wide range of frequencies.  For a cylindrical cavity and a 
homogeneous longitudinal magnetic field, $C=0.69$ for the lowest 
TM mode.  The form factors of the other modes are much smaller.  The 
resonant frequency of the lowest TM mode of a cylindrical cavity
is $f$=115 MHz $\left( {1m\over R}\right)$ where $R$ is the radius
of the cavity.  Since $10^{-6}\hbox{\ eV} = 2\pi$ (242 MHz), a
large cylindrical cavity is convenient for searching the low
frequency end of the range of interest.  To extend the search
to high frequencies without sacrifice in volume, one may
power-combine many identical cavities which fill up the
available volume inside a magnet's bore \cite{rsci,hag}.  This 
method allows one to maintain $C=0(1)$ at high frequencies, albeit 
at the cost of increasing engineering complexity as the number of 
cavities increases.

Pilot experiments were carried out at Brookhaven National
Laboratory \cite{RBF} and at the University of Florida \cite{UF}.  
These experiments used relatively small magnets and hence the 
limits they placed on the local axion dark matter density are not 
severe.  However they developed the various aspects of the cavity 
detection technique and demonstrated its feasibility.

Second generation experiments are presently under way at Lawrence 
Livermore National Laboratory (LLNL) \cite{LLNL} and at Kyoto
University \cite{Kyoto}.  The LLNL experiment is similar in concept 
to the UF pilot experiment but uses a much larger magnet 
($B_0^2 V = 12 T^2m^3$).  It is well engineered and runs with a near 
100\% duty cycle \cite{NIM}.  The results from its first year of 
running are reported in ref. \cite{LLNL98}.  The exclusion plot is 
shown in Fig.~2.  By definition, 
$g_{a\gamma\gamma} = {\alpha\over\pi} {g_\gamma\over f_a}$.
The limits shown assume that the local halo density, estimated to be
$7.5~10^{-25}~$g/cm$^3$ \cite{Gates}, is entirely in axions. 
The experiment has ruled out the hypothesis that 100\% of the local 
halo density is in KSVZ axions with mass in the range shown.  Since 
then, the frequency range 500-800 MHz ($2.1 \le m_a \le 3.3 \mu$eV)
has been searched but the results have not been published yet.  Up 
till now, the experiment has used a single cavity with a variety of 
dielectric rods and metal posts.  However, a four-cavity array will
soon be used to search higher frequencies.  Ultimately, the LLNL 
experiment will cover the mass range $1.3~\cdot 10^{-6}~$eV to 
$13~\cdot 10^{-6}~$eV at KSVZ sensitivity or better (see below).  

A development project is under way to equip the LLNL detector with 
SQUID microwave receivers.  These would replace the HEMT receivers
presently in use.  The HEMT receivers have noise temperature 
$T_n \sim 3~K$ \cite{Bradley}.  It appears that $T_n \ltwid 0.3~K$ 
will be reached with the SQUIDs \cite{Clarke}.  To take advantage of 
such low electronic noise temperatures, the experiment will have to 
be equipped with a dilution refrigerator.  Also bucking coils must 
be installed to cancel the static magnetic field at the location of 
the SQUID.  When this development project is completed, the LLNL 
detector will have sufficient sensitivity to detect DFSZ axions at 
even a fraction of the local halo density. 

The Kyoto experiment exploits resonant $a\to \gamma$ conversion in 
a cavity permeated by a large static magnetic field, as do the other 
experiments, but uses a beam of Rydberg atoms to count the photons 
from $a\to \gamma$ conversion \cite{Kyoto}.  Single photon counting 
constitutes a dramatic improvement in microwave detection sensitivity.  
With HEMT amplifiers one needs to have thousands of $a\to \gamma$ 
conversions per second and integrate for about 100~sec to find a 
signal in the noise.  With single photon counting, a few $a\to \gamma$ 
conversions suffice in principle.  To build a beam of Rydberg atoms 
capable of single photon counting is a considerable achievement in 
itself.  In addition, a dilution refrigerator is necessary to cool 
the cavity down to a temperature ($\sim 10$~mK) where the thermal 
photon background is negligible.  The projected sensitivity of the 
Kyoto experiment is sufficient to detect DFSZ axions at even a 
fraction of the local halo density.

\section{Other axion searches}

There are a number of other techniques which have been used to search 
for very weakly coupled (so-called 'invisible') axions.  Although 
these searches have not ruled out parameter space that is not also
presently ruled out by the astrophysical limits described above, they 
do provide completely independent constraints.

\subsection{Solar axion searches}

The conversion of axions to photons in a magnetic field can be used 
to look for solar axions too \cite{ps,KVB,Laz,Min}.  The flux of 
solar axions on Earth is   ${7.4\cdot 10^{11} \over 
{\rm sec}~{\rm cm}^2} ({g_\gamma \over 0.36})^2 ({m_a \over {\rm eV}})^2$ 
from the Primakoff conversion of thermal photons in the sun \cite{KVB}.  
The actual flux may be larger because other processes, such as Compton-like 
scattering, contribute if the axion has an appreciable coupling to
the electron.  At any rate the flux is huge compared to what can be 
produced by man-made processes on Earth and it is cost free.  Solar 
axions have a broad spectrum of energies of order the temperature in 
the solar core, from one to a few keV.  

Since the magnetic field is homogeneous on the length scale set by the 
axion de Broglie wavelength, the final photon is colinear with the initial 
axion.  The photon and axion also have the same energy assuming the
magnetic field is time-independent.  The $a \rightarrow \gamma$ conversion 
probability is \cite{ps,KVB}
\begin{equation}
p = {1 \over 4} ({\alpha g_\gamma \over \pi f_a})^2 (B_{0\perp} L F(q))^2
\label{prob}
\end{equation}
if $B_{0\perp} (z) = B_{0\perp} b(z)$ is the magnetic field transverse 
to the direction of the colinear axion and photon, $z$ is the coordinate 
along this direction, $L$ is the depth over which the magnetic field 
extends and $F(q)$ is the form factor
\begin{equation}
F(q) = {1 \over L} \mid \int_0^L dz e^{iqz} b(z) \mid
\end{equation}
where $q = k_\gamma - k_a = E_a - \sqrt{E_a^2 - m_a^2} 
\simeq {m_a^2 \over 2E_a}$ is the momentum transfer.  If the magnetic 
field is homogeneous ($b=1$), then
\begin{eqnarray}
F(q) &=& {2 \over qL} \mid sin{qL \over 2} \mid \nonumber\\
&\simeq& 1 ~~~~~~~~{\rm for}~~~~qL \ll 1~~~\ .
\label{ff}
\end{eqnarray}
For $qL \gg 1$, the conversion probability goes as $sin^2({qL \over 2})$ 
because the axion and photons oscillate into each other back and forth.
The form factor $F(q)$ can be improved by filling the conversion region 
with a gas whose pressure is adjusted in such a way that the plasma 
frequency, which acts as an effective mass for the photon, equals the 
axion mass \cite{KVB}.

Multiplying the flux times the conversion probability, one obtains the 
event rate:
\begin{equation}
{{\rm events} \over {\rm time}} \simeq
{200 \over {\rm day}} ~{V L \over {\rm meter}^4} ~F(q)^2 
({B_{0\perp} \over 8 {\rm Tesla}})^2 ({m_a \over {\rm eV}})^4 ~\ .
\end{equation}
The final state photons are soft x-rays which may be detected with good 
efficiency.  There are radioactive backgrounds to worry about however.

The above type of detector is usually referred to as an axion helioscope.
If a signal is found due to axions or familons, the detector immediately 
becomes a marvelous new tool for the study of the solar interior.  
Experiments were carried out at Brookhaven National Lab. \cite{Laz} and 
more recently at the University of Tokyo \cite{Min}.  The BNL experiment 
used a stationary Isabelle dipole magnet whose aperture was directed 
towards the sun at sunset.  The total exposure time was of order 
15 minutes. The Tokyo experiment uses a superconducting dipole magnet 
mounted on a altazimuth which tracks the sun.  The 95\% confidence level
upper limit \cite{Min} based on a few days of data taking is  
$g_{a\gamma\gamma} \equiv \alpha g_\gamma/\pi f_a \leq 
6.0 \cdot 10^{-10} {\rm GeV}^{-1}$ for $m_a \leq 0.03$ eV.

F. Avignone and his collaborators \cite{Avi} have exploited a different 
method to search for solar axions, namely the coherent Primakoff 
conversion of axions to photons in a crystal lattice.  When the 
incident angle fulfills the Bragg condition for a given crystalline 
plane, the rate is enhanced.  As the crystal detector turns with 
the Earth relative to the Sun's direction, a characteristic diurnal 
temporal pattern is produced. Using 1.94 kg.yr of data from a Ge
detector in Sierra Grande, Argentina, the bound $g_{a\gamma\gamma} <
2.7\cdot 10^{-9}$ GeV$^{-1}$ was obtained, independent of axion mass 
up to approximately 1 keV.     

S. Moriyama \cite{Moriyama} proposed looking for monochromatic 
axions emitted in the deexcitation of $^{54}{\rm Fe}$ in the sun.
$^{54}{\rm Fe}$ has an M1 transition, between the first excited 
state and the ground state, with excitation energy 14.4 keV.  The 
monochromatic axions can be resonantly absorbed by the same nucleus 
in the laboratory because the axions are Doppler broadened due to 
the thermal motion of the axion emitter in the Sun.  An experiment
of this type was carried out by M. Kr\v cmar et al. \cite{Krcmar}. 

\subsection{Laser experiments}

Eqs.(\ref{prob},\ref{ff}) give the conversion probability 
in a static magnetic field of an axion to a photon of the same energy.
The polarization of the photon is parallel to the component of the 
magnetic field transverse to the direction of motion.  The inverse 
process, conversion of such a photon to an axion, occurs with the 
same probability $p$, of course.  K. van Bibber et al. \cite{shine} 
proposed a 'shining light through walls' experiment in which a laser 
beam is passed through a long dipole magnet like those used for 
high-energy physics accelerators.  In the field of the magnet, a 
few of the photons convert to axions.  Another dipole magnet is set 
up in line with the first, behind a wall.  Since the axions go through 
the wall unimpeded, this setup allows one to 'shine light through 
the wall.'  An experiment of this type was carried out by the RBF 
collaboration \cite{Ruoso}.  Compared with a solar axion search, it 
has the advantage of greater control over experimental parameters.  
But the signal is much smaller because one pays twice the price of 
the very small axion-photon conversion rate.

Other types of laser experiments were proposed \cite{Maiani} in which 
one looks at the effect of the axion on the propagation of light 
through a magnetic field.  If the photon beam is linearly polarized 
and the polarization direction is at an angle to the direction of the
magnetic field, the plane of polarization turns because the component 
of light polarized parallel to the magnetic field gets depleted whereas 
the perpendicular component does not.  There is an additional effect 
of birefringence because the component of light polarized parallel to 
the magnetic field mixes with the axion and hence moves more slowly 
than in vacuo.  Birefringence affects the ellipticity of the polarization 
as the light travels on.  The birefringence associated with the axion 
is considerably smaller than that due to the box diagram in QED, i.e. 
an electron running in a loop with four external photon lines.  The 
polarization rotation and the birefringence effects were searched for 
by the RBF collaboration \cite{Cameron}, which boosted these effects 
by passing the laser beam hundreds of times in an optical cavity within 
the magnet.  For more recent work, see ref. \cite{Zav}.

\subsection{A telescope search}

The axion decays to two photons at the rate: 
\begin{eqnarray}
\Gamma(a \rightarrow 2 \gamma) &=& g_\gamma^2 {\alpha^2 \over 64 \pi^3}
{m_a^3 \over f_a^2} \nonumber\\
&=& {g_\gamma^2 \over 6.8~10^{24} {\rm sec}} 
\left({m_a \over {\rm eV}}\right)^5 \ .
\end{eqnarray}
For axions in the 10 eV mass range, the decay rate is comparable to the 
age of the universe.  The dominant contribution to the cosmological energy 
density of such axions is thermal production \cite{KT}.  The energy density 
in thermal axions is proportional to the axion mass and becomes equal to 
the critical energy density at a mass of order 100 eV, the exact value 
depending on how many particle species annihilate after the axion decoupled.  
One can search for relic axions by looking for the monochromatic photons 
from their decay.  Such photons arrive to us from all directions but 
preferentially from large agglomerations of mass, such as clusters of 
galaxies.  The relative width ${\Delta \lambda \over \lambda}$ of the 
photon line from axion decay in galactic clusters is of order the 
virial velocity there, i.e $10^{-2}$.  By subtracting the spectrum 
of light from a galactic cluster from that of the nightsky 'off cluster', 
one can subtract some of the background.  The latter is dominated by 
lines in the spectrum of airglow \cite{tsar}.  A search of this 
type was carried out at Kitt Peak National Laboratory by the TSAR 
collaboration \cite{tsar}.  They placed the limit 
$g_{a\gamma\gamma} \leq 10^{-10} {\rm GeV}^{-1}$ in the range 
$3 \leq m_a \leq 8$ eV (3100-8300 \AA).  

\section{Macroscopic forces mediated by axions}

J. Moody and F. Wilczek \cite{Moody} analyzed the apparent deviations 
from the $1/r^2$ gravitational force law due to the exchange of virtual 
axions.  The coupling of the axion to a spin 1/2 fermion $f$ has the general 
form:
\begin{equation}
{\cal L}_{aff} = g_f {m_f \over v} a \overline f (i \gamma_5 + \theta_f) f
\label{CPv}
\end{equation}
when allowance is made for the fact that CP is violated in the electroweak 
interactions.  $g_f$ is of order one, whereas $\theta_f$ is of order 
$\overline \theta$, which is of order $10^{-17}$ \cite{EG,GR} in the
Standard Model.  The second term in Eq. (\ref{CPv}) produces a coupling 
of the axion field to the mass density of a macroscopic collection of
non-relativisitc fermions, whereas the first term produces a coupling 
of the axion field to the spin density of that macroscopic body.  The 
first type of coupling was called 'monopole', the second 'dipole'.  
The axion mediated forces between two macroscopic bodies therefore 
fall into three categories: monopole-monopole, monopole-dipole, and 
dipole-dipole.  The monopole-monopole force is suppressed by two powers 
of $\theta_f$ and therefore very small.  The dipole-dipole force has 
a very large background from ordinary magnetic forces.  This background
can be suppressed by using superconducting shields but not well enough 
for the axion mediated contribution to be detected.  The monopole-dipole 
is the least difficult to detect.  It is suppressed by only one factor 
of $\theta_f$ and it can be modulated by rotating the spin polarized body.
Experiments of this type have been carried out \cite{Youdin}.  
Unfortunately, their sensitivity is many orders of magnitude 
short of what is required to see the effect.

\end{document}